\documentstyle[12pt,bezier]{article}
\newfont{\mathea}{msam10 scaled\magstep0}
\newfont{\matheb}{msbm10 scaled 1095}
\newfont{\tmpEins}{cmsy10 scaled 2074}
\newfont{\tmpZwei}{cmsy10 scaled 1095}
\newfont{\tmpDrei}{cmsy10 scaled 1000}
\newfont{\tmpVier}{cmsy5 scaled 1000}
\newfont{\tmpFuenf}{msbm7 scaled\magstep0}

\def\Bbb#1{\mathchoice{\mbox{\matheb #1}}{\mbox{\matheb #1}}%
 {\mbox{\tmpFuenf #1}}{\mbox{\tmpFuenf #1}}}

\def\restriction{\mathchoice{
 \mbox{\unitlength1cm\begin{picture}(.2,.4)%
  \bezier{5}(.07,.3)(.1,.27)(.13,.24)%
  \put(.07,.35){\line(0,-1){.5}}\end{picture}}}{
 \mbox{\unitlength1cm\begin{picture}(.2,.4)%
  \bezier{5}(.07,.3)(.1,.27)(.13,.24)%
  \put(.07,.35){\line(0,-1){.5}}\end{picture}}}{
 \mbox{\mathea\symbol{22}}}{
 \mbox{\mathea\symbol{22}}}}

\def\dach#1#2{\mbox{$\mathop{\vbox{\ialign{%
  $##\crcr\hfil #1 \hfil$\crcr}}}\limits^{\scriptscriptstyle #2}$}}

\def\rnzs{\dach{\rho_2}{\mbox{$\scriptscriptstyle\kern-.7mm0$}}\kern-1.2mm'}

\def\Subset{\mbox{$\subset\kern-.5mm\subset$}}
\newcommand{\LI}{\mbox{{\rm L$^{\kern-.15em\raise.2ex\hbox{\scriptsize 1}}$}}}

\def\Ldummy{\left.\bgroup}
\def\Rdummy{\egroup^{\rule{0mm}{1.4mm}}\right.}
\def\LA{\left\langle\bgroup}
\def\RA{\egroup^{\rule{0mm}{1.4mm}}\right\rangle_{\cal A}^{}}
\def\LR{\left(\bgroup}
\def\RR{\egroup^{\rule{0mm}{1.4mm}}\right)}
\def\LG{\left\{\bgroup}
\def\RG{\egroup^{\rule{0mm}{1.4mm}}\right\}}

\def\Wort#1{\mbox{{\rm #1\kern.1em}}}

\def\lfac#1#2{\vcenter{\hbox{$#1\kern-.2em\raise-.6ex\hbox{\Large{/}}%
 \kern-.2em\raise-1.2ex\hbox{$#2$}$}}}

\def\gin{\mbox{\tmpZwei\symbol{91}\kern-1.4mm\rule{.2mm}{1.85mm}\kern1.4mm}}
\def\gni{\mbox{\tmpZwei\symbol{92}\kern-1.4mm\rule[.15mm]{.2mm}{1.85mm}%
  \kern1.4mm}}

\def\EINS{{\mathop{1\kern-.25em\mbox{{\rm{\small l}}}}}}

\begin{document}

\LARGE Generalized Eigenvectors for Resonances in the Friedrichs Model and
Their Associated Gamov Vectors

\normalsize

\vspace{1cm}

Hellmut Baumg\"artel

\vspace{3mm}

\begin{abstract}
A Gelfand triplet for the Hamiltonian $H$ of the Friedrichs model on $\Bbb{R}$
with multiplicity space ${\cal K}, \dim\,{\cal K}<\infty$
is constructed such that exactly the resonances (poles of the inverse of the 
Liv\v{s}ic-matrix) are (generalized) eigenvalues of $H$. The corresponding
eigen(anti-)linearforms are calculated explicitly. Using the wave matrices for
the wave (M\"oller) operators the corresponding eigen(anti-)linearforms on the 
Schwartz space ${\cal S}$ for the unperturbed Hamiltonian $H_{0}$ are also 
calculated. It turns out that they are of pure Dirac type and can be characterized
by their corresponding Gamov vector $\lambda\rightarrow k/(\zeta_{0}-
\lambda)^{-1},\zeta_{0}$ resonance, $k\in{\cal K}$, which is uniquely determined
by restriction of ${\cal S}$ to ${\cal S}\cap {\cal H}^{2}_{+}$, where 
${\cal H}^{2}_{+}$ denotes the Hardy space of the upper half plane. 
Simultaneously this restriction yields a truncation of the generalized evolution
to the well-known decay semigroup for $t\geq 0$ of the Toeplitz type on
${\cal H}^{2}_{+}$. That is: exactly those pre-Gamov vectors
$\lambda\rightarrow k/(\zeta -\lambda)^{-1},\,\zeta$ from the lower half plane,
$k\in{\cal K}$, have an extension to a generalized eigenvector of $H$ if
$\zeta$ is a resonance and if $k$ is from that subspace of ${\cal K}$
which is uniquely determined by its corresponding Dirac type anti-linearform.
\end{abstract}

\vspace{3mm}


\vspace{3mm}


\section{Introduction}
In quantum scattering systems bumps in cross sections often can be described by 
expressions like $\lambda\rightarrow c((\lambda-\lambda_{0})^{2}
+(\frac{\Gamma}{2})^{2})^{-1}$, where $\lambda_{0}$ is the {\em resonance energy},
$\Gamma/2$ the {\em halfwidth}, called Breit-Wigner formulas (see e.g.
Bohm [1, pp. 428 - 429]). Sometimes, if the scattering matrix is analytically
continuable into the lower half plane $\Bbb{C}_{-}$, these bumps can be connected
with complex poles $\lambda_{0}-i\frac{\Gamma}{2}$ of the scattering matrix
in $\Bbb{C}_{-}$. Then $c((\lambda-\lambda_{0})-i\frac{\Gamma}{2})^{-1}$
is called the Breit-Wigner amplitude, if the pole is of first order (see e.g.
[1, pp. 428 - 429]). These poles are called {\em resonances} (see e.g.
Br\"andas/Elander [2], Albeverio/Ferreira/Streit [3]).

The basic idea is that these points should coincide with eigenvalues for
generalized eigenvectors of the evolution which is determined by the
Hamiltonian $H$ of the scattering system. Obviously this (first) problem
cannot be solved within the Hilbert space ${\cal H}$, it requires extension
techniques, e.g. the use of Gelfand triplets.

A further (second) problem is to establish a rigorous mathematical framework
to derive modified associated states, also corresponding
to resonances as eigenvectors, but of a {\em truncated evolution}, such that
the eigenvectors satisfy the exponential decay law. These vectors are called
{\em Gamov vectors} in the literature (see e.g. Gamov [4], Bohm/Gadella [5],
Bohm/Harshman [6] and further references therein). An obvious suggestion
is that also this problem has to be solved by techniques beyond the 
Hilbert space. Such an approach was presented by Bohm/Gadella and others by 
using Gelfand triplets (Rigged Hilbert Spaces (RHS) in their terminology)
on Hardy subspaces of ${\cal H}_{0}$, the Hilbert space of the unperturbed
Hamiltonian $H_{0}$ of the scattering system (see [5, 6], Bohm/Maxson/
Loewe/Gadella [7] and papers quoted therein).

Originally, the theory of Gelfand triplets (see e.g. Gelfand/Wilenkin [8],
see also Baumg\"artel [9]) was developed for selfadjoint operators to
generalize eigenvector expansions also for the absolutely continuous
spectrum. For this purpose the occurence of complex eigenvalues is only
a nuisance.

In this paper it is shown that for the finite-dimensional Friedrichs model
the first problem can be solved rigorously by the Gelfand triplet approach,
i.e. the construction of a triplet is presented such that exactly the
resonances are eigenvalues of the extended Hamiltonian. The corresponding
(generalized) eigenvectors are calculated explicitly (a slightly modified
triplet was already considered in Baumg\"artel [10]). This result confirms
the {\em basic idea} mentioned above.

On the other hand, recently it turned out that to solve the second problem
the use of the triplet approach is not indispensable. On the contrary, the
Gamov vectors can be identified as vectors in the Hilbert space ${\cal H}_{0}$
resp. ${\cal H}$, more precisely, they are eigenvectors of the
{\em decay semigroup} for $t\geq 0$, which is of Toeplitz type and which
can be defined by a truncation of the quantum evolution. This insight came
into the light and was supported by analogies in the Lax-Phillips scattering
theory. This approach has been promoted and emphasized by Strauss [11] (see
also Eisenberg/Horwitz/Strauss [12]).

However, if one adopts this point of view then a third problem arises: One
has to point out the connection between the generalized eigenvector (the
solution of the first problem) and the corresponding Gamov vector, i.e. one
has to determine the selection principle which selects the {\em right}
Gamov vector from the whole collection of all pre-Gamov vectors (eigenvectors
of the decay semigroup). Also this problem is solved in this paper: Exactly
those eigenvectors of the decay semigroup have extensions to a generalized 
eigenvector if the eigenvalue is a resonance and which belong to a
distinguished subeigenspace, which is calculated explicitly. Vice versa, the
restriction of the generalized eigenvector (for $H_{0}$)
which is an eigen(anti-)linearform 
on the Schwartz space of pure Dirac type 
to the Hardy subspace for the upper half plane
$\Bbb{C}_{+}$ is (via the Paley-Wiener theorem) characterized by a vector from
this Hardy space. This vector is the
Gamov vector corresponding to the generalized eigenvector.

\section{Preliminaries}
\subsection{Basic objects of the Friedrichs model}
In the following we collect the concepts and denotations for the 
finite-dimensional Friedrichs model on $\Bbb{R}$. Let
${\cal H}_{0}:=L^{2}(\Bbb{R},{\cal K},d\lambda)$,
where
${\cal K}$
denotes a multiplicity Hilbert space,
$\dim{\cal K}<\infty$.
Further let
${\cal E}$
be a finite-dimensional Hilbert space, $\dim{\cal E}=:N$
and put
${\cal H}:={\cal H}_{0}\oplus {\cal E}$.
The projection onto ${\cal E}$ is denotated by $P_{\cal E}.\, H_{0}$
is a selfadjoint operator on
${\cal H}$
with reducing projection $P_{\cal E}$,
where
$H_{0}\restriction {\cal H}_{0}$
is the multiplication operator on
${\cal H}_{0}$.
The selfadjoint operator $H$ on ${\cal H}$ is given by a perturbation of
${\cal H}_{0}$ as
\[
H:= H_{0}+\Gamma +\Gamma^{\ast},
\]
where $\Gamma$ denotes a partial isometry on ${\cal H}$ with the properties
\[
\Gamma^{\ast}\Gamma=P_{\cal E},\quad \Gamma\Gamma^{\ast}\leq P^{\bot}_{\cal E}
:=\EINS - P_{\cal E}.
\]
The operator function
\[
L_{\pm}(z):=(z-H_{0})P_{\cal E}-\Gamma^{\ast}(z-H_{0})^{-1}\Gamma, \quad
z\in\Bbb{C}_{\pm},
\]
the so-called Liv\v{s}ic-matrix, is decisive in the following. One has
$L_{\pm}(z)\restriction{\cal E}\in{\cal L}({\cal E})$
is holomorphic on $\Bbb{C}_{\pm}$. For brevity, if there is no danger of
confusion, we write
$L_{\pm}(z)$
instead of
$L_{\pm}(z)\restriction {\cal E}$.
Further we need the so-called partial resolvent
$P_{\cal E}(z-H)^{-1}P_{\cal E}$.
It turns out that
\[
L_{\pm}(z)\cdot P_{\cal E}(z-H)^{-1}P_{\cal E}=
P_{\cal E}(z-H))^{-1}P_{\cal E}\cdot L_{\pm}(z)=
P_{\cal E},\quad z\in\Bbb{C}_{\pm},
\]
(see e.g. Baumg\"artel [10]),
that is
\[
P_{\cal E}(z-H)^{-1}P_{\cal E}\restriction{\cal E}=
(L_{\pm}(z)\restriction{\cal E})^{-1},\quad z\in\Bbb{C}_{\pm},
\]
and this equation shows that
$(L_{\pm}(z)\restriction{\cal E})^{-1}\in{\cal L}({\cal E})$
is holomorphic on $\Bbb{C}_{\pm}$.

\vspace{3mm}

For ${\cal H}\ni x:=f+e,\, f\in{\cal H}_{0},\,e\in{\cal E}$ one has
$\Gamma x=\Gamma e,\,\Gamma^{\ast}x=\Gamma^{\ast}f$.
Therefore
\[
(\Gamma e)(\lambda)=M(\lambda)e,\quad {\cal E}\ni\Gamma^{\ast}f=
\int_{-\infty}^{\infty}M(\lambda)^{\ast}f(\lambda)d\lambda,
\]
where
$\lambda\rightarrow M(\lambda)\in{\cal L}({\cal E}\rightarrow{\cal K})$
is a.e. defined on $\Bbb{R}$.

\vspace{2mm}

{\em Assumption} 1: $M(\cdot)$ is a Schwartz function, i.e.
$M(\cdot)\in{\cal S}({\cal L}({\cal E}\rightarrow{\cal K})).$

\vspace{2mm}

For example, this implies
\[
\int_{-\infty}^{\infty}\Vert M(\lambda)^{\ast}M(\lambda)\Vert^{2}_
{2,{\cal E}}d\lambda <\infty,\quad
\int_{-\infty}^{\infty}\Vert M(\lambda)^{\ast}M(\lambda)\Vert_{2,{\cal E}}
d\lambda <\infty,
\]
where
$\Vert\cdot\Vert_{2,{\cal E}}$
denotes the Hilbert-Schmidt norm on ${\cal E}$.
Obviously one has
\begin{equation}
\Gamma^{\ast}(z-H_{0})^{-1}\Gamma\restriction {\cal E} =\int_{-\infty}^{\infty}
\frac{M(\lambda)^{\ast}M(\lambda)}{z-\lambda}d\lambda,\quad z\in\Bbb{C}_{\pm}.
\end{equation}
Therefore
s-$\lim_{\epsilon\rightarrow +0}\Gamma^{\ast}
(\lambda \pm i\epsilon -H_{0})^{-1}\Gamma$
exists on $\Bbb{R}$, hence also
$L_{\pm}(\lambda):=\\$s-$\lim_{\epsilon\rightarrow +0}L_{\pm}(\lambda\pm i\epsilon)$
exists and it is infinitely differentiable and polynomially bounded. From (1)
we obtain
\[
\frac{\Gamma^{\ast}E_{0}(d\lambda)\Gamma}{d\lambda}\restriction {\cal E}=
M(\lambda)^{\ast}M(\lambda),\quad \lambda\in\Bbb{R},
\]
where $E_{0}(\cdot)$ denotes the spectral measure of $H_{0}$
on ${\cal H}_{0}$.

\vspace{2mm}

{\em Assumption} 2: $H$ has no eigenvalues. This is equivalent to
$\det L_{+}(\lambda)\neq 0$ for all $\lambda\in\Bbb{R}$ (see e.g.
Baumg\"artel [10]).

\vspace{2mm}

Then $L_{+}(\lambda)^{-1}$ exists for all $\lambda\in\Bbb{R}$,
it is infinitely differentiable and
$\sup_{\lambda}\Vert L_{+}(\lambda)^{-1}\Vert_{\cal E}<\infty.$
Furthermore we have
\begin{equation}
\mbox{s-}\lim_{\epsilon\rightarrow +0}
P_{\cal E}(\lambda\pm i\epsilon - H)^{-1}P_{\cal E}\restriction {\cal E}
=(L_{\pm}(\lambda)\restriction {\cal E})^{-1},\quad \lambda\in\Bbb{R}.
\end{equation}
$H$ has no singular continuous spectrum. From (2) we obtain
\[
\frac{P_{\cal E}E(d\lambda)P_{\cal E}}{d\lambda}\restriction {\cal E}=
\frac{1}{2\pi i}(L_{-}(\lambda)^{-1}-L_{+}(\lambda)^{-1})
=L_{\pm}(\lambda)^{-1}M(\lambda)^{\ast}M(\lambda)L_{\mp}(\lambda)^{-1},
\quad \lambda \in\Bbb{R},
\]
where $E(\cdot)$ denotes the spectral measure of $H$.

\subsection{Wave operators and wave matrices}
Since $\Gamma +\Gamma^{\ast}$ is a finite-dimensional perturbation the
wave operators
$W_{\pm}=W_{\pm}(H,H_{0}):=$s-$\lim_{t\rightarrow\pm\infty}
 e^{itH}e^{-itH_{0}}P_{\cal E}^{\bot}$
exist, they are isometric from ${\cal H}_{0}$ onto ${\cal H}$.
Furthermore,
$W_{\pm}^{\ast}=W_{\pm}(H_{0},H)=$s-$\lim_{t\rightarrow\pm\infty}
e^{itH_{0}}e^{-itH}$.

\vspace{2mm}

In the following we rewrite the wave operators as limits of operator spectral 
integrals. We refer to Baumg\"artel/Wollenberg [15] for details on
operator spectral integrals, where this theory is presented. We use also results
of Baumg\"artel [13] (see also [14]). Here we mention only the following facts:
If
$\mu\rightarrow t(\mu):=\sum_{j=1}^{m}\chi_{\Delta_{j}}(\mu)t_{j},\,
t_{j}\in{\cal H}_{0}$, 
is a step function then the spectral integral
$\int_{-\infty}^{\infty}E_{0}(d\mu)t(\mu)$
is given by
\[
\int_{-\infty}^{\infty}E_{0}(d\mu)t(\mu)=
\sum_{j=1}^{m}\int_{-\infty}^{\infty}E_{0}(d\mu)\chi_{\Delta_{j}}(\mu)t_{j}=
\sum_{j=1}^{m}\int_{-\infty}^{\infty}\chi_{\Delta_{j}}(\mu)E_{0}(d\mu)t_{j}=
\sum_{j=1}^{m}E_{0}(\Delta_{j})t_{j}.
\]
The spectral integral
$\int_{-\infty}^{\infty}E_{0}(d\mu)x(\mu)$
for a more general function
$\mu\rightarrow x(\mu)\in{\cal H}_{0}$
exists if
\[
\int_{-\infty}^{\infty}\frac{(x(\lambda),E_{0}(d\mu)x(\lambda))}
{d\mu}\mid_{\mu=\lambda}d\lambda <\infty.
\]
Note that
$\frac{(g,E_{0}(d\mu)g)}{d\mu}$
exists a.e. on $\Bbb{R}$ for all $g\in{\cal H}_{0}$ because the spectral
measure $E_{0}(\cdot)$ is absolutely continuous.

\vspace{2mm}

Now put
${\cal H}_{E_{0}}:=\mbox{clo\,spa}(E_{0}(\Delta)f,\,f\in\Gamma{\cal E})$
and
${\cal H}_{E}:=\mbox{clo\,spa}(E(\Delta)e,\,e\in{\cal E}).$
It is not hard to see that
${\cal H}_{E_{0}}$ and ${\cal H}_{E}$
have natural spectral representations w.r.t. $E_{0}(\cdot),\, E(\cdot)$,
rspectively, which are explicitly given by spectral integrals:
\begin{equation}
{\cal H}_{E_{0}}\ni x=\int_{-\infty}^{\infty}E_{0}(d\mu)\Gamma f(\mu),\quad
{\cal H}_{E}\ni y=\int_{-\infty}^{\infty}E(d\lambda)g(\lambda),
\end{equation}
where $\mu\rightarrow f(\mu)\in{\cal E},\,\lambda\rightarrow g(\lambda)\in
{\cal E}$
are vector functions with values in ${\cal E}$ such that the integrals
(3) exist. Note that
$\int_{-\infty}^{\infty}E_{0}(d\mu)\Gamma f(\mu)$
exists iff
$\int_{-\infty}^{\infty}\Vert M(\mu)f(\mu)\Vert^{2}_{\cal K}d\mu<\infty$,
i.e. iff the function
$\mu\rightarrow M(\mu)f(\mu)$
is an element of
${\cal H}_{0}$.
The integral
$\int_{-\infty}^{\infty}E(d\lambda)g(\lambda)$
exists iff
$\int_{-\infty}^{\infty}\Vert M(\lambda)L_{+}(\lambda)^{-1}g(\lambda)
\Vert^{2}_{\cal K}d\lambda<\infty,$
i.e. iff the function
$\lambda\rightarrow M(\lambda)L_{+}(\lambda)^{-1}g(\lambda)$
is an element of ${\cal H}_{0}$.
The function $f(\cdot)$ is called the {\em representer} of $x$ and
$g(\cdot)$ the {\em representer} of $y$ w.r.t. the corresponding spectral
representation.

Note further that
\[
\left(\int_{-\infty}^{\infty}E_{0}(d\mu)\Gamma f(\mu)\right)(\lambda)=
\left(\Gamma f(\lambda)\right)(\lambda)=M(\lambda)f(\lambda)
\]
and
\[
{\cal H}_{0}\ominus{\cal H}_{E_{0}}=\{f\in{\cal H}_{0}:
M(\lambda)^{\ast}f(\lambda)=0\quad\mbox{a.e. on}\,\Bbb{R}\}.
\]
The wave operators 
$W_{\pm},\,W_{\pm}^{\ast}$
can be written as strong limits of certain spectral integrals (see [13]):
\begin{equation}
{\cal H}_{0}\ni f\rightarrow W_{\pm}f=
\mbox{s-}\lim_{\epsilon\rightarrow +0}\int_{-\infty}^{\infty}
E(d\lambda)\left(\EINS-\Gamma^{\ast}R_{0}(\lambda\pm i\epsilon)\right)f,
\end{equation}
\begin{equation}
{\cal H}\ni g\rightarrow W_{\pm}^{\ast}g=
\mbox{s-}\lim_{\epsilon\rightarrow +0}\int_{-\infty}^{\infty}
E_{0}(d\lambda)\left(\EINS+(\Gamma+\Gamma^{\ast})R(\lambda\pm i\epsilon)\right)g,
\end{equation}
where 
$R_{0}(z):=(z-H_{0})^{-1},\,R(z):=(z-H)^{-1}$
denote the resolvent of
$H_{0},\,H$ on ${\cal H}_{0},\,{\cal H}$,
respectively. From (4) we get immediately
\begin{equation}
W_{\pm}f=f,\quad f\in{\cal H}_{0}\ominus{\cal H}_{E_{0}}.
\end{equation}
$W_{\pm}$ on ${\cal H}_{E_{0}}$ and
$W_{\pm}^{\ast}$ on ${\cal H}_{E}$
can be calculated explicitly.

\vspace{3mm}

LEMMA 1. {\em The wave operators are given by the following expressions}:
\begin{equation}
W_{\pm}\left(\int_{-\infty}^{\infty}E_{0}(d\mu)\Gamma f(\mu)\right)=
\int_{-\infty}^{\infty}E(d\lambda)L_{\pm}(\lambda)f(\lambda),
\end{equation}
\begin{equation}
W_{\pm}^{\ast}\left(\int_{-\infty}^{\infty}E(d\lambda)g(\lambda)\right)=
\int_{-\infty}^{\infty}E_{0}(d\lambda)\Gamma L_{\pm}(\lambda)^{-1}g(\lambda).
\end{equation}

\vspace{2mm}

Proof. (7): First we calculate
$W_{\pm}(\Gamma e),\,e\in{\cal E}.$ From (4) we obtain
\begin{eqnarray*}
W_{\pm}(\Gamma e) &=&
\mbox{s-}\lim_{\epsilon\rightarrow +0}
\int_{-\infty}^{\infty}E(d\lambda)(\Gamma e-\Gamma^{\ast}R_{0}(\lambda\pm i
\epsilon)\Gamma e) \\
&=& \mbox{s-}\lim_{\epsilon\rightarrow +0}\int_{-\infty}^{\infty}E(d\lambda)
\left(\Gamma e+L_{\pm}(\lambda\pm i\epsilon)e-
((\lambda\pm i\epsilon)-H_{0})e\right) \\
&=& \mbox{s-}\lim_{\epsilon\rightarrow +0}\int_{-\infty}^{\infty}E(d\lambda)
\left(L_{\pm}(\lambda\pm i\epsilon)e-\lambda e\mp i\epsilon e+
H_{0}e+\Gamma e\right) \\
&=& \mbox{s-}\lim_{\epsilon\rightarrow +0}\int_{-\infty}^{\infty}E(d\lambda)
\left(L_{\pm}(\lambda\pm i\epsilon)e+(H-\lambda)e\right),
\end{eqnarray*}
but
$\int_{-\infty}^{\infty}E(d\lambda)(H-\lambda)e=0,$ i.e.
\[
W_{\pm}(\Gamma e)=\mbox{s-}\lim_{\epsilon\rightarrow +0}\int_{-\infty}^{\infty}
E(d\lambda)L_{\pm}(\lambda\pm i\epsilon)e.
\]
Now the spectral integral
$\int_{-\infty}^{\infty}E(d\lambda)L_{\pm}(\lambda)e$
exists and it turns out by straightforward calculation that one can interchange
s-lim and integral, i.e. finally we have
\[
W_{\pm}(\Gamma e)=\int_{-\infty}^{\infty}E(d\lambda)L_{\pm}(\lambda)e.
\]
Straightforward extension to the spectral integrals
$\int_{-\infty}^{\infty}E_{0}(d\mu)\Gamma f(\mu)$
yields (7).

(8) Correspondingly, first we calculate
$W_{\pm}^{\ast}e.$
According to (5) we have
\begin{eqnarray*}
W_{\pm}^{\ast}e &=&
\mbox{s-}\lim_{\epsilon\rightarrow +0}\int_{-\infty}^{\infty}E_{0}(d\lambda)
P_{\cal E}^{\bot}(\Gamma+\Gamma^{\ast})R(\lambda\pm i\epsilon)e \\
&=& \mbox{s-}\lim_{\epsilon\rightarrow +0}\int_{-\infty}^{\infty}E_{0}(d\lambda)
\Gamma P_{\cal E}R(\lambda\pm i\epsilon)P_{\cal E}e \\
&=& \mbox{s-}\lim_{\epsilon\rightarrow +0}\int_{-\infty}^{\infty}E_{0}(d\lambda)
\Gamma L_{\pm}(\lambda\pm i\epsilon)^{-1}e.
\end{eqnarray*}
Again, the spectral integral
$\int_{-\infty}^{\infty}E_{0}(d\lambda)\Gamma L_{\pm}(\lambda)^{-1}e$
exists and we can interchange s-lim and integral, i.e. we arrive at
\[
W_{\pm}^{\ast}e=\int_{-\infty}^{\infty}E_{0}(d\lambda)\Gamma
L_{\pm}(\lambda)^{-1}e.
\]
Extension to the spectral integrals
$\int_{-\infty}^{\infty}E(d\lambda)g(\lambda)$
gives (8).$\quad \Box$

\vspace{2mm}

Therefore
$W_{\pm}({\cal H}_{E_{0}})={\cal H}_{E}$
and
$W_{\pm}({\cal H}_{0}\ominus{\cal H}_{E_{0}})={\cal H}\ominus{\cal H}_{E}.$
Using (6) we get
${\cal H}_{0}\ominus{\cal H}_{E{0}}={\cal H}\ominus{\cal H}_{E}$.
Note that this is compatible with
${\cal E}\subset {\cal H}_{E}$.
Thus, the wave operators act nontrivially only on
${\cal H}_{E_{0}},\,{\cal H}_{E}.$

Lemma 1 says: if
$\lambda\rightarrow f(\lambda)$
is the representer of
$x\in{\cal H}_{E_{0}}$
w.r.t. $E_{0}$
then the representer of
$W_{\pm}x\in{\cal H}_{E}$
w.r.t. $E$ is given by
$\lambda\rightarrow L_{\pm}(\lambda)f(\lambda).$
Conversely, if
$\lambda\rightarrow g(\lambda)$
is the representer of
$y\in{\cal H}_{E}$
w.r.t. $E$
then the representer of
$W_{\pm}^{\ast}y\in{\cal H}_{E_{0}}$
w.r.t. $E_{0}$
is given by
$\lambda\rightarrow L_{\pm}(\lambda)^{-1}g(\lambda).$

In general, operator functions with these properties are called the
{\em wave matrices} of $W_{\pm},\,W_{\pm}^{\ast}$ w.r.t. given fixed 
spectral representations (see Baumg\"artel/Wollenberg [15, p. 177] for
these concepts). Note that wave matrices are well-defined only if the
spectral representations are fixed.

\vspace{3mm}

Lemma 2. {\em The wave matrices of}
$W_{\pm},\,W_{\pm}^{\ast}$
{\em wr.t. the natural spectral representations in}
${\cal H}_{E_{0}},\,{\cal H}_{E}$
{\em are given by}
\[
W_{\pm}(\lambda)=L_{\pm}(\lambda),\quad
W_{\pm}^{\ast}(\lambda)=L_{\pm}(\lambda)^{-1},\quad \lambda\in\Bbb{R}.
\]
Note that in the natural spectral representation of
${\cal H}_{E_{0}}$
the vectors
$\Gamma e,\,e\in{\cal E}$
are considered in some sense as "constants", whereas the corresponding function as
a function in  ${\cal H}_{0}$ w.r.t.the usual ${\cal K}$-representation is given
by
$\lambda\rightarrow (\Gamma e)(\lambda)=M(\lambda)e.$

As is well-known (see e.g. Baumg\"artel/Wollenberg [15, p. 398 ff.]) the
scattering matrix
$S_{\cal K}(\lambda):=(W_{+}^{\ast}W_{-})(\lambda)$
in the usual ${\cal K}$-representation of
${\cal H}_{0}={\cal H}_{E_{0}}\oplus ({\cal H}_{0}\ominus{\cal H}_{E_{0}})$
is given by
\begin{equation}
S_{\cal K}(\lambda)=\EINS_{\cal K}-2\pi i M(\lambda)L_{+}(\lambda)^{-1}
M(\lambda)^{\ast},\quad \lambda\in\Bbb{R}.
\end{equation}

\vspace{3mm}

LEMMA 3. {\em On}
${\cal H}_{E_{0}}$ {\em and w.r.t. the natural spectral representation of}
${\cal H}_{E_{0}}$ {\em the scattering matrix}
$S_{\cal E}(\cdot)$
{\em is given by}
\begin{equation}
S_{\cal E}(\lambda)=L_{+}(\lambda)^{-1}L_{-}(\lambda)=
L_{+}(\lambda)^{-1}L_{+}(\lambda)^{\ast}.
\end{equation}
{\em This means if}
$f\in{\cal H}_{E_{0}}$
{\em and}
$\tilde{f}(\cdot)$
{\em is its representer w.r.t.} $E_{0}$, {\em i.e.}
$f(\lambda)=M(\lambda)\tilde{f}(\lambda)$
{\em then}
$S_{\cal E}(\lambda)\tilde{f}(\lambda)$
{\em is the} $E_{0}$-{\em representer of}
$Sf$,
{\em where}
$(Sf)(\lambda)=S_{\cal K}(\lambda)f(\lambda).$

\vspace{2mm}

Proof. We have to prove that
$S_{\cal K}(\lambda)M(\lambda)\tilde{f}(\lambda)=M(\lambda)
S_{\cal E}(\lambda)\tilde{f}(\lambda).$
But this is obvious because of
\begin{equation}
M(\lambda)L_{+}(\lambda)^{-1}L_{-}(\lambda)=
(\EINS_{\cal K}-2\pi i M(\lambda)L_{+}(\lambda)^{-1}M(\lambda)^{\ast})M(\lambda)=
S_{\cal K}(\lambda)M(\lambda),
\end{equation}
$\Box$

\vspace{3mm}

REMARK 1. In the following we restrict the consideration to the case that
$\Gamma{\cal E}$
is generating for $H_{0}$ and ${\cal E}$ is generating for $H$, i.e. we 
assume
${\cal H}_{E}={\cal H}$ and ${\cal H}_{E_{0}}={\cal H}_{0}.$
This implies
$\dim\,{\cal E}=\dim\,{\cal K}$. Moreover, the operator function
$\lambda\rightarrow M(\lambda)\in{\cal L}({\cal E}\rightarrow {\cal K})$
is then invertible for all $\lambda,\,M(\lambda)^{-1}\in
{\cal L}({\cal K}\rightarrow{\cal E}).$

\section{Gelfand Triplets}
\subsection{The Schwartz space triplet on ${\cal H}_{0}$ and its transformation
to ${\cal H}$}
By ${\cal S}$ we denote the space of all Schwartz functions
$\lambda\rightarrow s(\lambda)\in{\cal K}$
with values in ${\cal K}$. The canonical norms on ${\cal S}$
are denoted by $\Vert\cdot\Vert_{\sigma}$,
where $\sigma$ labels these norms. 
${\cal S}\subset{\cal H}_{0}$
is dense in ${\cal H}_{0}$ w.r.t. the Hilbert space norm of ${\cal H}_{0}$.
The space of all continuous anti-linearforms on ${\cal S}$
is denoted by ${\cal S}^{\times}.$ Then
\[
{\cal S}\subset{\cal H}_{0}\subset{\cal S}^{\times}
\]
is a Gelfand triplet w.r.t. ${\cal H}_{0}$, the Schwartz space triplet.
The representer of $s$ in the $E_{0}$-representation is denoted by
$\tilde{s},\,s(\lambda)=M(\lambda)\tilde{s}(\lambda),\,
\lambda\rightarrow \tilde{s}(\lambda)\in{\cal E}.$

\vspace{2mm}

By the wave operator $W_{+}$ the Schwartz space triplet can be transformed
to a triplet w.r.t. ${\cal H}$. We put
${\cal D}:=W_{+}{\cal S}$
and equip ${\cal D}$ with the topology of ${\cal S}$. Thus we obtain the triplet
\begin{equation}
{\cal D}\subset{\cal H}\subset{\cal D}^{\times}.
\end{equation}
Note that
${\cal D}^{\times}=W_{+}^{\times}{\cal S}^{\times}$,
where ${\cal D}^{\times}\ni d^{\times}=W_{+}^{\times}s^{\times}$
is defined by
\[
\langle W_{+}^{\ast}d\mid s^{\times}\rangle=
\langle d\mid W_{+}^{\times}s^{\times}\rangle,\quad d\in{\cal D}.
\]

\vspace{3mm}

LEMMA 4. {\em The triplet} (12) {\em satisfies the following properties:}
\begin{itemize}
\item[(i)]
${\cal E}\subset{\cal D}$
{\em and}
${\cal E}=W_{+}{\cal T}$
{\em where}
${\cal T}:=\{f\in{\cal H}_{0}: f(\lambda)=M(\lambda)L_{+}(\lambda)^{-1}e,\,
e\in{\cal E}\}$
{\em is an} N-{\em dimensional subspace of}
${\cal H}_{0}$
{\em with}
${\cal T}\subset{\cal S}$,
\item[(ii)]
${\cal D}=\Phi\oplus{\cal E}$
{\em where}
$\Phi:=\{W_{+}s: s\in{\cal S}\cap({\cal H}_{0}\ominus{\cal T})\}=
P_{\cal E}^{\bot}{\cal D}\subset{\cal H}_{0}$,
\item[(iii)]
${\cal D}^{\times}=\Phi^{\times}\times{\cal E}$ 
{\em (cartesian product) where}
$\Phi^{\times}$
{\em is the space of all continuous anti-linearforms on}
$\Phi$,
\item[(iv)]
{\em if}
$d=\phi+e$
{\em and}
$d^{\times}=\{\phi^{\times},e^{\times}\}$
{\em then}
$\langle d\mid d^{\times}\rangle = \langle\phi\mid\phi^{\times}\rangle+
(e,e^{\times})_{\cal E}$.
\item[(v)]
$H_{0}\Phi\subseteq\Phi$ {\em and} $H{\cal D}\subseteq{\cal D}$.
\end{itemize}

\vspace{2mm}

Proof. (i)-(iv) are obvious because of Lemma 1. (v) is true because $H_{0}$
and $H$ act on the representers of elements in
$\Phi,\,{\cal D}$
by multiplication of the spectral parameters, respectively. $\Box$

\subsection{A modified Gelfand triplet}
Recall that
spec$(H_{0}\restriction{\cal E})$
is a finite set of (real) eigenvalues. Let
$(a,b)\subset\Bbb{R}$
be an open interval with
spec$(H_{0}\restriction{\cal E})\subset(a,b).$
Further let
$G_{0}\subset\Bbb{C}$
an (open) connected symmetric region (symmetric w.r.t. complex conjugation)
such that
$G_{0}\cap\Bbb{R}=(a,b)$.

\vspace{2mm}

{\em Assumption} 3. The operator function
$\Bbb{R}\ni\lambda\rightarrow M(\lambda)\in{\cal L}({\cal K}\rightarrow
{\cal E})$
has a holomorphic continuation into
$G_{0}.$

\vspace{2mm}

Then
$L_{+}(\cdot)$
is holomorphic in
$\Bbb{C}_{+}\cup G_{0}$
and
$L_{+}(\cdot)^{-1}$
is meromorphic there and even holomorphic in
$\Bbb{C}_{+}\cup (a,b).$

We introduce a modified Gelfand triplet: Recall first that the Schwartz functions
have the representation
$s(\lambda)=M(\lambda)L_{+}(\lambda)^{-1}x(\lambda),\,x(\lambda)\in{\cal E},$
where the representer in the $E_{0}$-representation is given by
$\tilde{s}(\lambda)=L_{+}(\lambda)^{-1}x(\lambda).$
Now let
${\cal S}_{0}\subset{\cal S}$
be the following submanifold of the Schwartz space:
\[
{\cal S}_{0}:=\{s\in{\cal S}: \lambda\rightarrow x(\lambda)\,
\mbox{is holomorphic continuable into}\, G_{0}\}.
\]
${\cal S}_{0}$
is dense in
${\cal S}$
w.r.t. the Schwartz topology. The (stronger) topology in
${\cal S}_{0}$
is defined by the collection of norms
\[
\Vert s_{0}\Vert_{\sigma,K}:=\Vert s_{0}\Vert_{\sigma}+
\sup_{z\in K\subset G_{0}}\Vert x(z)\Vert_{\cal E},
\]
where $K$ runs through all compact subsets of
$G_{0}.$ Then
\[
{\cal S}_{0}\subset {\cal H}_{0}\subset {\cal S}_{0}^{\times}
\]
is a modified Gelfand triplet w.r.t. ${\cal H}_{0}$.

The transformation of
${\cal S}_{0}$ to ${\cal H}$
is given, as before, by
${\cal D}_{0}:=W_{+}{\cal S}_{0}.$
Then
\[
{\cal D}_{0}\subset{\cal H}\subset{\cal D}_{0}^{\times}
\]
is a Gelfand triplet w.r.t. ${\cal H}$.
Similarly as in Lemma 4 we obtain 

\vspace{3mm}

LEMMA 5. {\em The modified Gelfand triplet satisfies the following properties}:
\begin{itemize}
\item[(i)]
${\cal E}\subset{\cal D}_{0},$
\item[(ii)]
${\cal D}_{0}=\Phi_{0}\oplus{\cal E}$,
{\em where}
$\Phi_{0}=P_{\cal E}^{\bot}{\cal D}_{0}$,
\item[(iii)]
${\cal D}_{0}^{\times}=\Phi_{0}^{\times}\times{\cal E}$
{\em and for}
$d_{0}=\phi_{0}+e,\,d_{0}^{\times}=\{\phi_{0}^{\times},e^{\times}\}$
{\em one has}
\[
\langle d_{0}\mid d_{0}^{\times}\rangle=\langle\phi_{0}\mid\phi_{0}^{\times}\rangle+
(e,e^{\times})_{\cal E}.
\]
\item[(iv)]
$H_{0}\Phi_{0}\subseteq\Phi_{0}$
{\em and}
$H{\cal D}_{0}\subseteq{\cal D}_{0}.$
\end{itemize}

\vspace{2mm}

Proof. (i) Since the functions
$x(\cdot)$
for the elements
$f\in{\cal T}$
are given by
$x(\lambda)=e$
for all $\lambda$, i.e. by constants, the condition of holomorphic
continuability is obviously satisfied. (ii)-(iv) are true because of
Lemma 4. $\Box$

\vspace{2mm}

REMARK 2. A simple example satisfying assumptions 1-3 is given for multiplicity
$N=1$, i.e. ${\cal E}=\Bbb{C}e_{0}$, then, according to Remark 1 one has also
${\cal K}=\Bbb{C}.$ Let
$\lambda_{0}\in\Bbb{R}$
be the eigenvalue of $H_{0},\, H_{0}e_{0}=\lambda_{0}e_{0}.$
Choose
$\Gamma e_{0}(\lambda):=e^{-\lambda^{2}/2}.$
Then
\[
\Gamma^{\ast}(z-H_{0})^{-1}\Gamma e_{0}=\int_{-\infty}^{\infty}
\frac{e^{-\lambda^{2}}}{z-\lambda}d\lambda\, e_{0}
\]
and
\[
L_{+}(z)=z-\lambda_{0}+\int_{-\infty}^{\infty}\frac{e^{-\lambda^{2}}}
{\lambda-z}d\lambda,
\]
where we have omitted the factor $e_{0}$.
Let $x_{0}\in\Bbb{R}$.
The calculation $z\rightarrow x_{0}+i0$ gives
\[
L_{+}(x_{0})=x_{0}-\lambda_{0}+i\pi e^{-x_{0}^{2}}+\int_{-\infty}^{\infty}
\frac{e^{-\lambda^{2}}}{\lambda-x_{0}}\,d\lambda,
\]
where the integral is Cauchy's mean value. This shows that $L_{+}(x_{0})=0$
is impossible because Cauchy's mean value is real. That is, the assumptions 1 and
2 are satisfied. Assumption 3 is satisfied because
$\lambda\rightarrow e^{-\lambda^{2}/2}$
is holomorphic in $\Bbb{C}$ hence
$z\rightarrow L_{+}(z)$
is also holomorphic in $\Bbb{C}$. The same is true for $L_{-}(\cdot)$.

\subsection{Resonances}
We define the concept {\em resonance} for the Friedrichs model satisfying 
Assumptions 1,2,3 as follows:

\vspace{2mm}

The point
$\zeta_{0}\in{\cal G}_{0}\cap\Bbb{C}_{-}$
is called a resonance if
$\det\,L_{+}(\zeta_{0})=0.$

\vspace{2mm}

In other words,
$\zeta_{0}$
is a resonance iff
$\zeta_{0}$
is a pole of
$L_{+}(\cdot)^{-1}$,
i.e. a pole of the analytic continuation of the partial resolvent into
$G_{0}\cap\Bbb{C}_{-}.$
From Lemma 3 we obtain: a point
$\zeta_{0}\in G_{0}\cap\Bbb{C}_{-}$
is a pole of
$L_{+}(\cdot)^{-1}$
iff it is a pole of
$S_{\cal K}(\cdot)$
resp. of
$S_{\cal E}(\cdot).$

\section{Results}
The first result (Theorem 1) says that exactly the resonances are 
eigenvalues of the extended Hamiltonian $H^{\times}$ w.r.t. the modified
Gelfand triplet for ${\cal H}$, if for the corresponding eigenvectors
a certain analyticity condition is required.

\vspace{3mm}

THEOREM 1. {\em The point}
$\zeta_{0}\in G_{0}\cap\Bbb{C}_{-}$
{\em is an eigenvalue of the extended Hamiltonian}
$H^{\times}$
{\em w.r.t. the Gelfand triplet}
${\cal D}_{0}\subset{\cal H}\subset{\cal D}_{0}^{\times}$
{\em with eigenanti-linearform}
$d_{0}^{\times}:=\{\phi_{0}^{\times}(\zeta_{0},e_{0}),e_{0}\}$
{\em satisfying the eigenvalue equation}
$H^{\times}d_{0}^{\times}=\zeta_{0}d_{0}^{\times},$
{\em where}
$\phi_{0}^{\times}(\zeta_{0},e_{0})$
{\em is the analytic continuation into}
$G_{0}\cap\Bbb{C}_{-}$
{\em of a holomorphic vector anti-linearform}
$\phi_{0}^{\times}(z,e_{0})$
{\em in}
$\Bbb{C}_{+}$
{\em iff}
$\zeta_{0}$
{\em is a resonance. The anti-linearform}
$\Bbb{C}_{+}\ni z\rightarrow \phi_{0}^{\times}(z,e)$
{\em is given by}
\[
\langle\phi\mid\phi_{0}^{\times}(z,e)\rangle:=
(\phi, (z-H_{0})^{-1}\Gamma e)_{{\cal H}_{0}},\quad \phi\in\Phi_{0},\,
z\in\Bbb{C}_{+},
\]
{\em and}
$e_{0}$
{\em satisfies}
$L_{+}(\zeta_{0})e_{0}=0,$
{\em i.e.}
$e_{0}\in\ker\,L_{+}(\zeta_{0}).$
{\em That is, the (generalized) eigenspace of}
$\zeta_{0}$
{\em is} q-{\em dimensional, where} q {\em is the geometric multiplicity
of the eigenvalue} 0 {\em of}
$L_{+}(\zeta_{0})$.

\vspace{3mm}

The second result (Theorem 2) concerns the structure of the corresponding
eigenanti-linearform 
$s_{0}^{\times}$
of $H_{0}^{\times}$
w.r.t. the modified Schwartz space triplet. This anti-linearform is given by
\[
s_{0}^{\times}(\zeta_{0},e_{0})=(W_{+}^{\ast})^{\times}
d_{0}^{\times}(\zeta_{0},e_{0}).
\]
It turns out that
$s_{0}^{\times}$
is an anti-linearform on
${\cal S}_{0}$
of a pure Dirac type w.r.t. the point
$\overline{\zeta_{0}}$
and there is a very simple transformation formula from
$e_{0}$
to the corresponding vector
$k_{0}\in{\cal K}$.

\vspace{3mm}

THEOREM 2. {\em The eigenanti-linearform}
$s_{0}^{\times}$
{\em of}
$H_{0}^{\times}$
{\em w.r.t. the Gelfand triplet}
${\cal S}_{0}\subset{\cal H}_{0}\subset{\cal S}_{0}^{\times},$
{\em associated to}
$d_{0}^{\times}$
{\em by}
$s_{0}^{\times}:=(W_{+}^{\ast})^{\times}d_{0}^{\times}$
{\em is given by}
\[
\langle s\mid s_{0}^{\times}(\zeta_{0},e_{0})\rangle=
2\pi i(s(\overline{\zeta_{0}}),k_{0})_{\cal K},\quad s\in{\cal S}_{0},
\]
{\em where}
$k_{0}:=M(\zeta_{0})e_{0}.$

\vspace{2mm}

The third result (Corollary 3) connects the eigenanti-linearform
$s_{0}^{\times}(\zeta_{0},e_{0})$
with a corresponding Gamov vector which is uniquely determined by
$s_{0}^{\times}.$

Recall that {\em pre-Gamov vectors} are considered (in this paper) as the
eigenvectors of the {\em truncated evolution}
$t\rightarrow Q_{+}e^{-itH_{0}}\restriction {\cal H}^{2}_{+},\,t\geq 0$,
where
${\cal H}^{2}_{+}\subset{\cal H}_{0}$
is the Hardy subspace for
$\Bbb{C}_{+}$
and
$Q_{+}$
the projection onto this Hardy subspace. The truncated evolution is a strongly
continuous contractive semigroup on
${\cal H}^{2}_{+}$
of the Toeplitz type (see e.g. Strauss [11]). As is well-known, each point
$\zeta\in\Bbb{C}_{-}$
is an eigenvalue of the generator of this semigroup and the corresponding
eigenspace is given by
$\{f\in{\cal H}^{2}_{+}: f(\lambda):=k(\lambda-\zeta)^{-1},\,k\in{\cal K}\},$
i.e. the dimension of the eigenspace of
$\zeta$
coincides with
$\dim{\cal K}$.

Now the decisive question is which pre-Gamov vectors are connected with
eigenanti-linearforms of
$H_{0}^{\times}.$
The first answer is that one has to select the poles of
$L_{+}(\cdot)^{-1}$ resp. of $S_{\cal K}(\cdot).$
However it remains the question: which values of
$k\in{\cal K}$
have to be chosen such that the pre-Gamov vector given by $k$ is in fact
connected to an eigenanti-linearform of
$H_{0}^{\times}$.

Recall first that
${\cal S}_{0}\cap{\cal H}^{2}_{+}\subset{\cal H}^{2}_{+}$
is dense in
${\cal H}^{2}_{+}$
w.r.t. the Hilbert space norm of
${\cal H}^{2}_{+}.$
The mentioned connection is then simply given by restriction of
$s_{0}^{\times}$
to
${\cal S}_{0}\cap{\cal H}^{2}_{+}.$

\vspace{3mm}

COROLLARY 3. {\em The restricted eigenanti-linearform}
$s_{0}^{\times}\restriction {\cal S}_{0}\cap{\cal H}^{2}_{+}$
\[
{\cal S}_{0}\cap{\cal H}^{2}_{+}\ni s\rightarrow
2\pi i(s(\overline{\zeta_{0}}),k_{0})_{\cal K}
\]
{\em is even continuous w.r.t. the Hilbert space topology of}
${\cal H}^{2}_{+}$,
{\em i.e. it can be continuously extended onto}
$\mbox{clo}({\cal S}_{0}\cap{\cal H}^{2}_{+})={\cal H}^{2}_{+}.$
{\em That is,}
$s_{0}^{\times}\restriction{\cal H}^{2}_{0}$
{\em is realized by the}
${\cal H}^{2}_{+}$-{\em vector}\\
$k_{0}(\zeta_{0}-\lambda)^{-1}$
{\em via the relation}
\begin{equation}
2\pi i(s(\overline{\zeta_{0}}),k_{0})=
\int_{-\infty}^{\infty}\left(s(\lambda),\frac{k_{0}}{\zeta_{0}-\lambda}\right)_
{\cal K}d\lambda.
\end{equation}

\vspace{2mm}

Proof. (13) follows immediately from the Paley-Wiener theorem.$\quad \Box$

\vspace{2mm}

Corollary 3 means: the restriction on
${\cal H}^{2}_{+}$
of the eigenanti-linearform
$s_{0}^{\times}$,
which is the back transform
$s_{0}^{\times}=(W_{+}^{\ast})^{\times}d_{0}^{\times}$
of
$d_{0}^{\times}$,
associated to the resonance $\zeta_{0}$ and to the parameter vector
$e_{0}\in\ker\,L_{+}(\zeta_{0})$,
to the Hilbert space
${\cal H}_{0}$
resp. the corresponding Gelfand triplet yields the associated Gamov vector
$\lambda\rightarrow k_{0}(\zeta_{0}-\lambda)^{-1}$,
where
$k_{0}=M(\zeta_{0})e_{0}$.
Conversely, exactly the pre-Gamov vectors where
$\zeta_{0}$
is a resonance and
$k_{0}=M(\zeta_{0})$
with
$e_{0}\in\ker\,L_{+}(\zeta_{0})$
have an extension (or "continuation") to an eigenanti-linearform of the extended
Hamiltonian
$H^{\times}$
w.r.t. the Gelfand triplet
${\cal D}_{0}\subset{\cal H}\subset{\cal D}_{0}^{\times}$.
That is exactly these pre-Gamov vectors are true Gamov vectors.

\vspace{2mm}

The last result presents a simple partial answer to the question,
how the parameter space
$M(\zeta_{0})\ker\,L_{+}(\zeta_{0})$
can be derived from the Laurent expansion of the scattering matrix
$S_{\cal E}(\cdot)$
at
$\zeta_{0}$.

\vspace{3mm}

PROPOSITION 4. {\em If}
$\zeta_{0}$
{\em is a simple pole of}
$S_{\cal E}(\cdot)$
{\em then}
\begin{equation}
\ker\,L_{+}(\zeta_{0})=\mbox{ima}\{\mbox{Res}_{z=\zeta_{0}}S_{\cal E}(z)\}.
\end{equation}

\vspace{2mm}

Proof. An easy calculation gives
\[
\ker\,L_{+}(\zeta_{0})=\mbox{ima}\,L_{-1}=\mbox{ima}(L_{-1}L_{+}
(\overline{\zeta_{0}})^{\ast}),
\]
where
$L_{-1}=\mbox{Res}_{z=\zeta_{0}}L_{+}(z)^{-1}.$
This gives (14). Note that
$L_{+}(\overline{\zeta_{0}})^{\ast})^{-1}$
exists.$\quad\Box$

\vspace{2mm}

REMARK 3. The relation between the order $g$ of the pole $\zeta_{0}$
of $S_{\cal E}(\cdot)$ and
$q:=\dim\,\ker\,L_{+}(\zeta_{0})$
is complicated. If
$m\leq N=\dim{\cal E}$
is the algebraic multiplicity of the eigenvalue $0$ of
$L_{+}(\zeta_{0})$ and
$r,\,1\leq r\leq m,$
the order of the zero $\zeta_{0}$ of
$\det\,L_{+}(z)$, then in any case
$1\leq g\leq r$ (see e.g. [16] for details).

\section{Proofs}
\subsection{Proof of Theorem 1}
The eigenvalue equation for eigenvalues
$\zeta_{0}\in G_{0}\cap\Bbb{C}_{-}$
of
$H^{\times}$ w.r.t. the triplet
${\cal D}_{0}\subset{\cal H}\subset{\cal D}_{0}^{\times}$
reads
\[
\langle d\mid H^{\times}d_{0}^{\times}\rangle=
\langle d\mid\zeta_{0}d_{0}^{\times}\rangle,\quad d\in {\cal D}_{0},
\]
or
\[
\langle Hd\mid d_{0}^{\times}\rangle=
\langle\overline{\zeta_{0}}d\mid d_{0}^{\times}\rangle,\quad d\in{\cal D}_{0},
\]
where
$d=\phi+e,\,\phi\in\Phi_{0},\,e\in{\cal E},\,d_{0}^{\times}=
\{\phi_{0}^{\times},e_{0}\},\,\phi_{0}^{\times}\in\Phi_{0}^{\times},\,
e_{0}\in{\cal E}.$
This is equivalent with
\[
(H_{0}e-\overline{\zeta_{0}}e,e_{0})+
\langle\Gamma e\mid\phi_{0}^{\times}\rangle=
\langle\overline{\zeta_{0}}\phi-H_{0}\phi\mid\phi_{0}^{\times}\rangle-
(\Gamma^{\ast}\phi,e_{0}).
\]
Since $e$ and $\phi$ vary independently we obtain two equations:
\begin{equation}
((\overline{\zeta_{0}}-H_{0})e,e_{0})=\langle\Gamma e\mid\phi_{0}^{\times}\rangle,
\quad e\in{\cal E},
\end{equation}
and
\begin{equation}
\langle(\overline{\zeta_{0}}-H_{0})\phi\mid\phi_{0}^{\times}\rangle=
(\Gamma^{\ast}\phi,e_{0}),\quad \phi\in\Phi_{0}.
\end{equation}
$\phi_{0}^{\times}$
depends on
$\zeta_{0}$,
the possible eigenvalue (and on $e_{0}$). According to our analyticity condition for
$\phi_{0}^{\times}$
this anti-linearform is required to be the analytic continuation of a holomorphic
vector anti-linearform
$\Bbb{C}_{+}\ni z\rightarrow\phi_{0}^{\times}(z)$.
This means that equation (16) has to be valid also on
$\Bbb{C}_{+}$
and it is a vector anti-linearform there:
\begin{equation}
((\overline{z}-H_{0})\phi,\phi_{0}^{\times}(z))_{{\cal H}_{0}}=
(\Gamma^{\ast}\phi,e_{0})_{\cal E},\quad z\in\Bbb{C}_{+},\quad\phi\in\Phi_{0},
\end{equation}
or
\[
(\phi,(z-H_{0})\phi_{0}^{\times}(z))_{{\cal H}_{0}}=
(\phi,\Gamma e_{0})_{\cal E},\quad z\in\Bbb{C}_{+},\quad \phi\in\Phi_{0}.
\]
This means
$(z-H_{0})\phi_{0}^{\times}(z)=\Gamma e_{0}$
or
\[
\phi_{0}^{\times}(z)=(z-H_{0})^{-1}\Gamma e_{0},\quad z\in\Bbb{C}_{+}.
\]
Now we have to check that this anti-linearform on
$\Phi_{0}$
is analytically continuable into
$\Bbb{C}_{+}\cup G_{0}$
as a holomorphic anti-linearform according to the requirement in Theorem 1:

We have shown in Subsection 3.2 that the elements
$s\in{\cal S}_{0}$ have the representation
$s(\lambda)=M(
\lambda)L_{+}(\lambda)^{-1}x(\lambda)$,
where
$\lambda\rightarrow x(\lambda)\in{\cal E}.$
Then
$(W_{+}s)(\lambda)=x(\lambda)$
and the function
$x(\cdot)$
is holomorphic continuable into
$G_{0}$.
If
$\zeta\in\Bbb{C}_{+}$
we have
\begin{eqnarray*}
\langle\phi\mid\phi_{0}^{\times}(\zeta)\rangle &=&
(P_{\cal E}^{\bot}W_{+}s,(\zeta-H_{0})^{-1}\Gamma e_{0}) \\
&=& (W_{+}s,(\zeta-H_{0})^{-1}\Gamma e_{0}) \\
&=& \left(\int_{-\infty}^{\infty}E(d\lambda)x(\lambda),(\zeta-H_{0})^{-1}
\Gamma e_{0}\right) \\
&=& \int_{-\infty}^{\infty}\frac{
(E(d\lambda)x(\lambda),(\zeta-H_{0})^{-1}\Gamma e_{0})}{d\lambda}d\lambda.
\end{eqnarray*}
Since
$x(\lambda)=\sum_{j=1}^{N}x_{j}(\lambda)b_{j}$,
where the
$\{b_{j}\}_{j}$
form an orthonormal basis of
${\cal E}$,
we obtain
\[
\langle\phi\mid\phi_{0}^{\times}(\zeta)\rangle=
\sum_{j=1}^{N}\int_{-\infty}^{\infty}\overline{x_{j}(\lambda)}
\frac{(E(d\lambda)b_{j},R_{0}(\zeta)\Gamma e_{0})}{d\lambda}d\lambda,
\]
so that we have to calculate the expression
\[
\frac{(E(d\lambda)e,R_{0}(\zeta)\Gamma e_{0})}{d\lambda}
\]
for any $e\in{\cal E}$. This calculation starts with the identity
\[
(R(z)e,R_{0}(\zeta)\Gamma e_{0})=
(R_{0}(z)\Gamma L_{+}(z)^{-1}e,R_{0}(\zeta)\Gamma e_{0}),
\quad z,\zeta\in\Bbb{C}_{+},
\]
where for the calculation of the right hand side the explicit expression for the
resolvent $R(z)=(z-H)^{-1}$ is used. This implies
\[
(R(\mu\pm i0)e,R_{0}(\zeta)\Gamma e_{0})=
\frac{1}{\mu-\zeta}\left((R_{0}(\overline{\zeta})\Gamma L_{\pm}(\mu)^{-1}e,
\Gamma e_{0})-
(R_{0}(\mu\pm i0)\Gamma L_{\pm}(\mu)^{-1}e,\Gamma e_{0})\right).
\]
Using
\[
\frac{E(d\mu)}{d\mu}=\frac{1}{2\pi i}\left(R(\mu-i0)-R(\mu+i0)\right)
\]
finally after a lengthy but straightforward calculation we obtain
\begin{equation}
\frac{(E(d\mu)e,R_{0}(\zeta)\Gamma e_{0})}{d\mu}=
\frac{1}{\mu-\zeta}\left(L_{\pm}(\mu)^{-1}M(\mu)^{\ast}M(\mu)
L_{\mp}(\mu)^{\-1}e,(\zeta-\mu-L_{+}(\zeta))e_{0}\right).
\end{equation}
Inspection of (18) proves the assertion. Now we know that the anti-linearform
$\phi_{0}^{\times}(z)$
satisfies the equation (17) for
$z\in\Bbb{C}_{+}$.
Therefore
$\phi_{0}^{\times}(\zeta,e_{0})$
satisfies the equation (16) for all
$\zeta\in G_{0}\cup\Bbb{C}_{+}$ (where now we have taken into account the second
parameter $e_{0}$).
Since
$z\rightarrow\phi_{0}^{\times}(z,e_{0})$
is holomorphic in the whole region
$G_{0}\cup\Bbb{C}_{+}$
we consider the (second) equation (15) first on
$\Bbb{C}_{+}$.
Then it reads
\[
((\overline{z}-H_{0})e,e_{0})=\langle\Gamma e\mid\phi_{0}^{\times}(z,e_{0}\rangle=
(\Gamma e,(z-H_{0})^{-1}\Gamma e_{0})=(e,\Gamma^{\ast}(z-H_{0})^{-1}\Gamma e_{0})
\]
so that we have
\begin{equation}
(e,(z-H_{0})e_{0})-\langle\Gamma e\mid\phi_{0}^{\times}(z,e_{0})\rangle=
(e,L_{+}(z)e_{0}),\quad e\in{\cal E},\quad z\in\Bbb{C}_{+},
\end{equation}
and the equation (15) reads simply
$(e,L_{+}(z)e_{0})=0$
for all $e\in{\cal E}$
which obviously has no solution in
$\Bbb{C}_{+}\cup(a,b)$.
But by analytic continuation the identity (19) is true also in
$\Bbb{C}_{-}\cap G_{0}$.
That is, equation (15) is equivalent to
\begin{equation}
L_{+}(\zeta_{0})e_{0}=0,\quad \zeta_{0}\in\Bbb{C}_{-}\cap G_{0}.
\end{equation}
This means: equation (15) has a solution $\zeta_{0}$ with corresponding
parameter $e_{0}\in{\cal E}$
iff equation (20) is satisfied. Conversely, if $\zeta_{0}\in\Bbb{C}_{-}\cap G_{0}$
and $e_{0}\in{\cal E}$
satisfy equation (20) then
$\zeta_{0}$
is an eigenvalue of
$H^{\times}$
and
$d_{0}^{\times}:=\{\phi_{0}^{\times}(\zeta_{0},e_{0}),e_{0}\}$
is a corresponding eigenanti-linearform. The dimension of the eigenspace of
$\zeta_{0}$
is then
$\dim\,\ker L_{+}(\zeta_{0}).\quad \Box$

\subsection{Proof of Theorem 2}
To calculate
$s_{0}^{\times}(\zeta_{0},e_{0})=(W_{+}^{\ast})^{\times}d_{0}^{\times}$
with
$d_{0}^{\times}=\{\phi_{0}^{\times}(\zeta_{0},e_{0}),e_{0}\}$
first we consider
$\phi_{0}^{\times}$
again for
$z\in\Bbb{C}_{+}$
and calculate
$s_{0}^{\times}(z,e_{0})=(W_{+}^{\ast})^{\times}\{\phi_{0}^{\times}(z,e_{0}),e_{0}\}.$
Later on we consider the analytic continuation into
$G_{0}\cap\Bbb{C}_{-}.$
We start with
\begin{eqnarray*}
\langle s\mid s_{0}^{\times}(z,e_{0})\rangle &=&
\langle W_{+}s\mid d_{0}^{\times}(z,e_{0})\rangle \\
&=&\langle P_{\cal E}^{\bot}W_{+}s\mid\phi_{0}^{\times}(z,e_{0})\rangle +
(P_{\cal E}W_{+}s,e_{0}) \\
&=& (P_{\cal E}^{\bot}W_{+}s,(z-H_{0})^{-1}\Gamma e_{0})+
(P_{\cal E}W_{+}s,e_{0}) \\
&=& (W_{+}s,(z-H_{0})^{-1}\Gamma e_{0})+(s,W_{+}^{\ast}e_{0})
\end{eqnarray*}
We have
$W_{+}^{\ast}e_{0}=\int_{-\infty}^{\infty}E_{0}(d\lambda)\Gamma L_{+}(\lambda)^{-1}
e_{0}d\lambda$
and
$W_{+}s=\int_{-\infty}^{\infty}E(d\lambda)L_{+}(\lambda)\tilde{s}(\lambda)$,
where
$s=\int_{-\infty}^{\infty}E_{0}(d\lambda)\Gamma\tilde{s}(\lambda),$
i.e.
$\tilde{s}(\cdot)$
is the representer of $s$ w.r.t. the
$E_{0}$-representation,
$s(\lambda)=M(\lambda)\tilde{s}(\lambda).$
Then
\[
(W_{+}s,R_{0}(z)\Gamma e_{0})=
\int_{-\infty}^{\infty}\frac{(E(d\lambda)L_{+}(\lambda)\tilde{s}(\lambda),
R_{0}(z)\Gamma e_{0})}{d\lambda}\,d\lambda.
\]
Again we use (18) for the calculation of this expression and obtain
\[
(W_{+}s,R_{0}(z)\Gamma e_{0}) =
\]
\[
\int_{-\infty}^{\infty}\frac{1}{\mu-z}\left(
L_{-}(\mu)^{-1}M(\mu)^{\ast}M(\mu)L_{+}(\mu)^{-1}L_{+}(\mu)\tilde{s}(\mu),
(z-\mu-L_{+}(z))e_{0}\right)d\mu =
\]
\[
 -\int_{-\infty}^{\infty}(L_{-}(\mu)^{-1}M(\mu)^{\ast}s(\mu),e_{0})d\mu + 
\int_{-\infty}^{\infty}\frac{1}{z-\mu}\left(
L_{-}(\mu)^{-1}M(\mu)^{\ast}M(\mu)\tilde{s}(\mu),L_{+}(z)e_{0}\right)d\mu.
\]
Furthermore we have
\begin{eqnarray*}
(s,W_{+}^{\ast}e_{0}) &=&
\left(s,\int_{-\infty}^{\infty}E_{0}(d\lambda)\Gamma L_{+}(\lambda)^{-1}
e_{0}d\lambda\right) \\
&=& \int_{-\infty}^{\infty}\left(s(\lambda),M(\lambda)L_{+}(\lambda)^{-1}
e_{0}\right)_{\cal K}d\lambda \\
&=& \int_{-\infty}^{\infty}\left(L_{-}(\lambda)^{-1}M(\lambda)^{\ast}
s(\lambda),e_{0}\right)_{\cal E}d\lambda,
\end{eqnarray*}
so that we finally obtain
\[
\langle s\mid s_{0}^{\times}(z,e_{0})\rangle=
\left(\int_{-\infty}^{\infty}\frac{1}{\overline{z}-\mu}
L_{-}(\mu)^{-1}M(\mu)^{\ast}s(\mu)d\mu,L_{+}(z)e_{0}\right)_{\cal E}.
\]
For the analytic continuation from
$z\in\Bbb{C}_{+}$ into $\Bbb{C}_{+}\cup G_{0}$
we have to check the integral
\begin{equation}
\Psi_{-}(\overline{z}):=\int_{-\infty}^{\infty}
\frac{1}{\overline{z}-\mu}L_{-}(\mu)^{-1}M(\mu)^{\ast}s(\mu)d\mu.
\end{equation}
Since this integral is the left factor in the scalar product we
substitute for the moment $z\rightarrow\overline{z}$,
consider
\begin{equation}
\Psi_{-}(z):=\int_{-\infty}^{\infty}\frac{1}{z-\mu}L_{-}(\mu)^{-1}
M(\mu)^{\ast}s(\mu)d\mu,\quad z\in\Bbb{C}_{-},
\end{equation}
and check the continuation into $\Bbb{C}_{+}$.
Recall that
$z\rightarrow\Psi_{+}(z)$ for $z\in\Bbb{C}_{+}$ is defined by one and the same 
formula (22). Then we obtain for $z\in\Bbb{C}_{+}$
\begin{eqnarray*}
\Psi_{-}(z) &=&
\Psi_{+}(z)+2\pi i L_{-}(z)^{-1}M(\overline{z})^{\ast}s(z) \\
&=& \Psi_{+}(z)+2\pi i(L_{+}(\overline{z})^{-1})^{\ast}
M(\overline{z})^{\ast}s(z).
\end{eqnarray*}
Substituting again $z\rightarrow\overline{z}$,
i.e. now we have $\overline{z}\in\Bbb{C}_{+}$
and
$z\in\Bbb{C}_{-}$, we obtain
\[
(\Psi_{-}(\overline{z}),L_{+}(z)e_{0})=
(\Psi_{+}(\overline{z}),L_{+}(z)e_{0})+
2\pi i ((L_{+}(z)^{-1})^{\ast}M(z)^{\ast}s(\overline{z}),L_{+}(z)e_{0}),
\]
where $\Psi_{+}(\overline{z})$ is a holomorphic part such that the first
term vanishes for $z=\zeta_{0}$. Then we have
\[
\langle s\mid s_{0}^{\times}(z,e_{0})\rangle=
2\pi i (M(z)^{\ast}s(\overline{z}),L_{+}(z)^{-1}L_{+}(z)e_{0})+
(\Psi_{+}(\overline{z}),L_{+}(z)e_{0})
\]
and
\[
\langle s\mid s_{0}^{\times}(\zeta_{0},e_{0})\rangle=
2\pi i(M(\zeta_{0})^{\ast}s(\overline{\zeta_{0}}),e_{0})=
2\pi i(s(\overline{\zeta_{0}}),M(\zeta_{0})e_{0})_{\cal K},
\]
that is, the anti-linearform
$s_{0}^{\times}(\zeta_{0},e_{0})$
is of pure Dirac type w.r.t. the point
$\overline{\zeta_{0}}$
and the corresponding vector
$k_{0}\in{\cal K}$
with
\[
\langle s\mid s_{0}^{\times}(\zeta_{0},e_{0})\rangle=
2\pi i (s(\overline{\zeta_{0}}),k_{0})_{\cal K}
\]
is given by
\[
k_{0}:=M(\zeta_{0})e_{0}.
\]
This confirms the fact (which is known from the beginning) that the
subspace of the admissible vectors
$k\in{\cal K}$
has the dimension
$\dim\ker\,L_{+}(\zeta_{0})$, too. $\quad \Box$

\section{Acknowledgement}
It is a pleasure to thank Professor A. Bohm for discussions on the subject
at the 3rd International Workshop on Pseudo-Hermitean Hamiltonians in
Quantum Physics at Ko\c{c} University, Istanbul, June 20 - 22 and at
DESY Zeuthen, July 5, 2005.

\section{References}
\begin{enumerate}
\item Bohm, A.: \\
Quantum Mechanics, Springer Verlag Berlin 1979
\item Br\"andas, E. and Elander, N. (eds.): \\
Resonances, Lecture Notes in Physics 325, Springer Verlag Berlin 1989
\item Albeverio, S., Ferreira, J.C. and Streit, L.: \\
Resonances - models and phenomena, in: Lecture Notes inPhysics 211, \\
Springer Verlag Berlin 1984
\item Gamov, G.: \\
Zur Quantentheorie des Atomkerns, Z. Phys. 51, 204 - 212 (1928)
\item Bohm, A. and Gadella, M.:\\
Dirac Kets, Gamov Vectors and Gelfand Triplets, Lecture Notes in Physics 348,\\
Springer Verlag Berlin 1989
\item Bohm, A. and Harshman, N. L.:\\
Quantum Theory in the Rigged Hilbert Space - Irreversibility from Causality, in:\\
Irreversibility and Causality, Semigroups and Rigged Hilbert Spaces,\\
Lecture Notes in Physics 504, Springer Verlag Berlin 1998
\item Bohm, A., Maxson, S., Loewe, M. and Gadella, M.:\\
Quantum mechanical irreversibility, Physica A 236, 485 - 549 (1997)
\item Gelfand, I. M. and Wilenkin, N. J.:\\
Verallgemeinerte Funktionen (Distributionen) IV, Einige Anwendungen der 
harmonischen Analyse, Gelfandsche Raumtripel,\\
VEB Deutscher Verlag der Wissenschaften, Berlin 1964
\item Baumg\"artel, H:\\
Resonanzen und Gelfandsche Raumtripel, Math. Nachr. 72, 93 - 98 (1976)
\item Baumg\"artel, H.:\\
Resonances of Perturbed Selfadjoint Operators and their Eigenfunctionals,\\
Math. Nachr. 75, 133 - 151 (1976)
\item Strauss, Y.:\\
Resonances in the Rigged Hilbert Space and Lax-Phillips Scattering Teory,\\
Internat. J. of Theor. Phys. 42, 2285 - 2317 (2003)
\item Eisenberg, E., Horwitz, L. P. and Strauss, Y.:\\
The Lax-Phillips Semigroup of the Unstable Quantum System, in:\\
Irreversibility and Causality, Semigroups and Rigged Hilbert Spaces,\\ 
Lecture Notes in Physics 504,
Springer Verlag Berlin 1998
\item Baumg\"artel, H.:\\
Eine Bemerkung zur Theorie der Wellenoperatoren,\\
Math. Nachr. 42, 359 - 363 (1969)
\item Baumg\"artel, H.:\\
Integraldarstellungen der Wellenoperatoren von Streusystemen,\\
Mber. Dt. Akad. Wiss. 9, 169 - 174 (1967)
\item Baumg\"artel, H. and Wollenberg, M.:\\
Mathematical Scattering Theory,\\
Operator Theory: Advances and Applications Vol. 9,\\
Birkh\"auser Verlag Basel, Boston, Stuttgart 1983
\item Baumg\"artel, H.:\\
Analytic Perturbation Theory for Matrices and Operators,\\
Operator Theory: Advances and Applications Vol. 15,\\
Birkh\"auser Verlag Basel Boston Stuttgart 1985

\end{enumerate}

\end{document}